\documentclass[fleqn,usenatbib,useAMS]{mnras}
 \usepackage{graphicx}
 \usepackage{amsmath}
 \usepackage{amssymb}
 \usepackage[T1]{fontenc}
 \usepackage{ae,aecompl}
 \usepackage{txfonts}

 \title[Commensurabilities between ETNOs]
       {Commensurabilities between ETNOs: a Monte Carlo survey}

 \author[C. de la Fuente Marcos and R. de la Fuente Marcos]
        {C.~de~la~Fuente~Marcos\thanks{E-mail: carlosdlfmarcos@gmail.com}
         and
         R. de la Fuente Marcos \\
         Apartado de Correos 3413, E-28080 Madrid, Spain}
 \date{Accepted 2016 April 20.
       Received 2016 April 20;
       in original form 2016 March 14}
 \pubyear{2016}
 
\begin{document}
  \label{firstpage}
  \pagerange{\pageref{firstpage}--\pageref{lastpage}}
  \maketitle

  \begin{abstract}
     Many asteroids in the main and trans-Neptunian belts are trapped 
     in mean motion resonances with Jupiter and Neptune, respectively.
     As a side effect, they experience accidental commensurabilities
     among themselves. These commensurabilities define characteristic
     patterns that can be used to trace the source of the observed 
     resonant behaviour. Here, we explore systematically the existence 
     of commensurabilities between the known ETNOs using their 
     heliocentric and barycentric semimajor axes, their uncertainties, 
     and Monte Carlo techniques. We find that the commensurability 
     patterns present in the known ETNO population resemble those 
     found in the main and trans-Neptunian belts. Although based on 
     small number statistics, such patterns can only be properly 
     explained if most, if not all, of the known ETNOs are subjected 
     to the resonant gravitational perturbations of yet undetected 
     trans-Plutonian planets. We show explicitly that some of the 
     statistically significant commensurabilities are compatible with 
     the Planet Nine hypothesis; in particular, a number of objects
     may be trapped in the 5:3 and 3:1 mean motion resonances with a
     putative Planet Nine with semimajor axis $\sim$700~au.
  \end{abstract}


  \section{Introduction}
     In Trujillo \& Sheppard (2014), the extreme trans-Neptunian objects or ETNOs are defined as asteroids with heliocentric semimajor axis 
     greater than 150 au and perihelion greater than 30 au. There are at present 16 known ETNOs (see Tables \ref{helio} and \ref{bary} for 
     relevant data); they exhibit clustering in the values of their argument of perihelion (Trujillo \& Sheppard 2014), longitude of the 
     ascending node (Batygin \& Brown 2016), eccentricity and inclination (de la Fuente Marcos \& de la Fuente Marcos 2014, 2016). The 
     analysis in de la Fuente Marcos \& de la Fuente Marcos (2014) has shown that the clustering in eccentricity ($e$ about 0.81) can be due 
     to a selection effect, but this cannot be the case of the ones found in inclination ($i$ about 20\degr), longitude of the ascending 
     node ($\Omega$ about 134\degr), and argument of perihelion ($\omega$ about $-$26\degr). These patterns could be induced by the 
     gravitational perturbation from one (Trujillo \& Sheppard 2014; Gomes, Soares \& Brasser 2015; Batygin \& Brown 2016) or more (de la 
     Fuente Marcos \& de la Fuente Marcos 2014; de la Fuente Marcos, de la Fuente Marcos \& Aarseth 2015), yet to be discovered, 
     trans-Plutonian planets.

     Malhotra, Volk \& Wang (2016) have pointed out that the four ETNOs with the longest orbital periods (see Table \ref{bary}) have simple
     numerical relationships between periods. This implies that the ETNOs could be a highly structured, dynamically speaking, population. 
     The ETNOs are not massive enough to induce orbit-orbit coupling by themselves, but a hypothetical planet or planets could be the source
     of these accidental commensurabilities. Malhotra et al. (2016) used their analysis to predict the existence of a planet with semimajor
     axis $\sim$665~au that is only marginally compatible with the one discussed by Batygin \& Brown (2016) within the framework of their
     Planet Nine hypothesis; even if their proposed value for the semimajor axis is similar, $\sim$665~au versus $\sim$700~au, the preferred 
     values of $i$ and $\Omega$ in Malhotra et al. (2016) ---($i\sim18\degr, \Omega\sim281\degr$) or ($i\sim48\degr, \Omega\sim355\degr$)--- 
     are rather different from those in Batygin \& Brown (2016) ---($i\sim30\degr, \Omega\sim113\degr$).

     Accidental commensurabilities are a natural by-product of the existence of mean motion resonances in the Solar system (for the theory
     of orbital resonance, see e.g. Murray \& Dermott 1999). The orbital architecture of the main asteroid belt is mainly the result of 
     interior mean motion resonances with Jupiter (see e.g. Holman \& Murray 1996; Nesvorn\'y \& Morbidelli 1999); conversely, the one 
     observed in the trans-Neptunian belt is induced by exterior mean motion resonances with Neptune (see e.g. Gladman et al. 2012). In the 
     main asteroid belt, the dominant ones are located at 2.06~au (4:1), 2.5~au (3:1), 2.82~au (5:2), 2.95~au (7:3) and 3.27~au (2:1); 
     therefore, any asteroids trapped in these resonances are also nominally resonant among themselves even if these objects are not massive 
     enough to induce orbit-orbit coupling by themselves. For instance, the Alinda asteroids which are subjected to the 3:1 mean motion 
     resonance with Jupiter and the Griqua asteroids that occupy the 2:1 resonance are also in a mutual 3:2 near mean motion resonance. For 
     objects trapped in the dominant resonances, the values of the ratios of orbital periods are:
     $$
     \begin{array}{l}
        1.0699, 1.1670, 1.2172, 1.2487, 1.2818, 1.3369, 1.4959, \\ 1.6017, 1.7137, 2.0000\,.
     \end{array}
     $$
     This sequence represents a characteristic sample of accidental commensurabilities in the main asteroid belt. In the trans-Neptunian 
     belt, the dominant mean motion resonances with Neptune are located at 39.4~au (2:3), 42.3~au (3:5), 43.7~au (4:7), 47.8~au (1:2) and
     55.4~au (2:5); the associated ratios of orbital periods are:
     $$
     \begin{array}{l}
        1.0501, 1.1124, 1.1440, 1.1681, 1.2012, 1.2477, 1.3363, \\ 1.4274, 1.4988, 1.6673\,.
     \end{array}
     $$
     They define accidental commensurabilities beyond Neptune. Out of 10 values, as many as five could be common to both sequences.

     The commensurability analysis carried out in Malhotra et al. (2016) provides some support for the Planet Nine hypothesis and singles 
     out some resonant orbital relations, but it does not explore the subject of commensurabilities between ETNOs in depth. Here, we use the
     seminal idea presented in Malhotra et al. (2016) and the fact that very eccentric orbits are affected by mean motion resonances no 
     matter the value of the semimajor axis (see e.g. Gomes et al. 2005) to investigate this subject extensively. We use their heliocentric 
     and barycentric semimajor axes, their uncertainties, and Monte Carlo techniques to find out if the values of the ratios of orbital 
     periods of the ETNOs match those in the main and trans-Neptunian belts. This Letter is organized as follows. Section~2 presents the 
     data used in this study. Our Monte Carlo methodology is explained and applied in Section~3. Results are discussed in Section~4 and 
     conclusions are summarized in Section~5. 

  \section{Data: heliocentric versus barycentric orbital elements}
     Ignoring the mass of the object, expressing distances in astronomical units, times in sidereal years, and masses in Solar masses, 
     Kepler's Third Law applied to an object orbiting the Sun states that $P^2=a^3$, where $P$ is the period and $a$ is the semimajor axis
     of the orbit of the object. Given two objects ---the first one with $a_{i}$ and $P_{i}$, and the second one with $a_{j}$ and $P_{j}$, 
     $i{\neq}j$--- Kepler's Third Law is given by $(P_{j}/P_{i})^2=(a_{j}/a_{i})^3$. If $P_{j}/P_{i}$ can be written in the form of a ratio 
     of small integers then we have a mean motion resonance. In order to investigate the distribution of ratios of orbital periods, 
     $P_{j}/P_{i}$, for the ETNOs the values of the semimajor axes are needed. 

     The orbital solutions of the ETNOs are far less robust than those available for the best studied members of the main asteroid belt or 
     the trans-Neptunian belt; the values of the relative error in semimajor axis are several orders of magnitude larger in the case of the 
     ETNOs. Any study of commensurabilities between ETNOs must account for this fact; a probabilistic approach based on the values of their 
     uncertainties can be used to address this important issue. The values of the heliocentric semimajor axes and their uncertainties from 
     the Jet Propulsion Laboratory, JPL, Small-Body Database\footnote{http://ssd.jpl.nasa.gov/sbdb.cgi} and \textsc{Horizons} On-Line 
     Ephemeris System (Giorgini et al. 1996) for the 16 objects discussed in this Letter are given in Table \ref{helio}. Three objects 
     ---2003~SS$_{422}$, 2010~GB$_{174}$ and 2013~RF$_{98}$--- have relative uncertainties in $a$ greater than five per cent as their orbits 
     are still in need of further improvement. However, the heliocentric orbital elements may not be adequate for this study as the objects 
     discussed here are too distant from the Sun. In cases like this one, the use of barycentric orbital elements is perhaps more 
     appropriate (Todorovic-Juchnicwicz 1981; Malhotra et al. 2016).
%
%
      \begin{table}
        \centering
        \fontsize{8}{11pt}\selectfont
        \tabcolsep 0.35truecm
        \caption{Heliocentric semimajor axes and their uncertainties for the 16 objects discussed in this Letter. (Epoch: 2457400.5, 
                 2016-January-13.0 00:00:00.0 TDB. J2000.0 ecliptic and equinox. Source: JPL Small-Body Database. Data retrieved on 2016 
                 March 13.)
                }
        \begin{tabular}{cc}
          \hline
             Object          & $a$ (au) \\
          \hline
              (82158) 2001 FP$_{185}$ & 226.34477$\pm$0.32205  \\
              (90377) Sedna           & 507.56034$\pm$0.80866  \\
             (148209) 2000 CR$_{105}$ & 227.95133$\pm$0.53837  \\
             (445473) 2010 VZ$_{98}$  & 152.77938$\pm$0.15108  \\
             2002~GB$_{32}$           & 215.76214$\pm$0.66575  \\
             2003~HB$_{57}$           & 164.61811$\pm$0.56922  \\
             2003~SS$_{422}$          & 193.83280$\pm$48.28200 \\
             2004~VN$_{112}$          & 321.01987$\pm$1.08350  \\
             2005~RH$_{52}$           & 151.13760$\pm$0.18708  \\
             2007~TG$_{422}$          & 492.72774$\pm$3.60060  \\
             2007~VJ$_{305}$          & 188.33726$\pm$0.55010  \\
             2010~GB$_{174}$          & 371.11831$\pm$28.60100 \\
             2012~VP$_{113}$          & 259.30017$\pm$7.17910  \\
             2013~GP$_{136}$          & 152.49683$\pm$0.56214  \\
             2013~RF$_{98}$           & 309.07383$\pm$37.05700 \\
             2015~SO$_{20}$           & 162.70354$\pm$0.13024  \\
          \hline
        \end{tabular}
        \label{helio}
      \end{table}
%
%

     Table \ref{bary} shows the values of various orbital parameters referred to the barycentre of the Solar system for the known ETNOs as 
     well as relevant descriptive statistics; in this table, unphysical values are displayed for completeness. The statistics is only 
     slightly different from that of the heliocentric orbital elements. The barycentric mean value of $e$ of the known ETNOs amounts to 
     0.81$\pm$0.06, the one of $i$ is 20\degr$\pm$8\degr, that of $\Omega$ is 134\degr$\pm$72\degr, and for $\omega$ is 
     $-$26\degr$\pm$49\degr; these barycentric values coincide with the heliocentric ones. Considering the conventional limits to detect 
     statistical outliers ---OL, lower outlier limit (Q$_{1}-1.5$IQR), and OU, upper outlier limit (Q$_{3}+1.5$IQR) with Q$_{1}$, first 
     quartile, Q$_{3}$, third quartile, IQR, interquartile range--- we observe that both (90377) Sedna and 2012 VP$_{113}$ are statistical 
     outliers in terms of perihelion distance, $q$, when barycentric orbits are considered. Sedna and 2007~TG$_{422}$ are also outliers in 
     orbital period, and 2003~SS$_{422}$ is an outlier in terms of $\omega$ (see the $\omega^*$ column in Table \ref{bary}). 
      
%
%
      \begin{table*}
        \centering
        \fontsize{8}{11pt}\selectfont
        \tabcolsep 0.10truecm
        \caption{Barycentric orbital elements and parameters ---$q=a(1-e)$, $Q=a(1+e)$, $\varpi=\Omega+\omega$, $P$ is the orbital period, 
                 $\Omega^*$ and $\omega^*$ are $\Omega$ and $\omega$ in the interval ($-\pi$, $\pi$) instead of the regular (0, 2$\pi$)--- 
                 for the 16 objects discussed in this Letter. The statistical parameters are Q$_{1}$, first quartile, Q$_{3}$, third 
                 quartile, IQR, interquartile range, OL, lower outlier limit (Q$_{1}-1.5$IQR), and OU, upper outlier limit (Q$_{3}+1.5$IQR); 
                 see the text for additional details. (Epoch: 2457400.5, 2016-January-13.0 00:00:00.0 TDB. J2000.0 ecliptic and equinox. 
                 Source: JPL Small-Body Database. Data retrieved on 2016 March 13.)
                }
        \begin{tabular}{lccccccccccc}
          \hline
             Object             & $a$ (au)  & $e$     & $i$ (\degr) & $\Omega$ (\degr) & $\omega$ (\degr) & $\varpi$ (\degr) & $q$ (au) & 
                       $Q$ (au) & $P$ (yr)    & $\Omega^*$ (\degr) & $\omega^*$ (\degr) \\
          \hline
                          82158 & 215.98190 & 0.84141 & 30.80136    & 179.35894        &   6.88445        & 186.24339        & 34.25243 &
                      397.71138 &  3172.07225 &    179.35894       &     6.88445        \\
                          Sedna & 506.08871 & 0.84945 & 11.92856    & 144.40252        & 311.28568        &  95.68820        & 76.19098 &
                      935.98645 & 11377.76605 &    144.40252       &  $-$48.71432       \\
                         148209 & 221.97203 & 0.80122 & 22.75599    & 128.28590        & 316.68920        &  84.97510        & 44.12266 &
                      399.82140 &  3304.94630 &    128.28590       &  $-$43.31080       \\
                         445473 & 153.36247 & 0.77602 &  4.51049    & 117.39868        & 313.72536        &  71.12404        & 34.35049 &
                      272.37444 &  1897.99758 &    117.39868       &  $-$46.27464       \\
                2002 GB$_{32}$  & 206.51087 & 0.82887 & 14.19246    & 177.04398        &  37.04706        & 214.09104        & 35.33979 &
                      377.68194 &  2965.72847 &    177.04398       &     37.04706       \\
                2003 HB$_{57}$  & 159.66636 & 0.76138 & 15.50028    & 197.87107        &  10.82968        & 208.70076        & 38.09895 &
                      281.23378 &  2016.21650 & $-$162.12893      &     10.82968       \\
                2003 SS$_{422}$ & 197.89878 & 0.80078 & 16.78599    & 151.04696        & 209.92850        &   0.97547        & 39.42454 &
                      356.37302 &  2782.15748 &    151.04696      & $-$150.07150       \\
                2004 VN$_{112}$ & 327.43683 & 0.85548 & 25.54760    &  66.02281        & 326.99697        &  33.01978        & 47.32201 &
                      607.55165 &  5921.18077 &     66.02281      &  $-$33.00303       \\
                2005 RH$_{52}$  & 153.67965 & 0.74624 & 20.44581    & 306.11082        &  32.53836        & 338.64918        & 38.99709 &
                      268.36221 &  1903.88874 &  $-$53.88918      &     32.53836       \\
                2007 TG$_{422}$ & 502.04619 & 0.92916 & 18.59529    & 112.91074        & 285.68512        &  38.59586        & 35.56266 &
                      968.52973 & 11241.71413 &    112.91074      &  $-$74.31488      \\
                2007 VJ$_{305}$ & 192.10299 & 0.81684 & 11.98375    &  24.38239        & 338.33483        &   2.71722        & 35.18469 &
                      349.02128 &  2660.83651 &     24.38239      &  $-$21.66517      \\
                2010 GB$_{174}$ & 351.12787 & 0.86169 & 21.56245    & 130.71445        & 347.24510        & 117.95955        & 48.56288 &
                      653.69286 &  6575.29094 &    130.71445      &  $-$12.75490      \\
                2012 VP$_{113}$ & 263.16571 & 0.69436 & 24.05155    &  90.80392        & 293.54965        &  24.35357        & 80.43515 &
                      445.89627 &  4266.39413 &     90.80392      &  $-$66.45035      \\
                2013 GP$_{136}$ & 149.78679 & 0.72587 & 33.53903    & 210.72729        &  42.47811        & 253.20540        & 41.06080 &
                      258.51277 &  1832.00761 & $-$149.27271      &     42.47811      \\
                2013 RF$_{98}$  & 317.06911 & 0.88557 & 29.60062    &  67.53385        & 316.37520        &  23.90906        & 36.28242 &
                      597.85579 &  5642.19269 &     67.53385      &  $-$43.62480      \\
                2015 SO$_{20}$  & 164.90527 & 0.79885 & 23.41102    &  33.63385        & 354.82995        &  28.46379        & 33.17008 &
                      296.64046 &  2116.25900 &     33.63385      &   $-$5.17005      \\
          \hline
             Mean               & 255.17510 & 0.81083 & 20.32577    & 133.64051        & 221.52645        & 107.66696        & 43.64735 &
                      466.70284 &  4354.79057 &     66.14051      &  $-$25.97355      \\ 
             Std. dev.          & 116.48640 & 0.06105 &  7.72494    &  71.95064        & 140.15771        & 102.40844        & 14.31074 &
                      226.78035 &  3103.34874 &    105.71229      &     49.06286      \\
             Median             & 211.24639 & 0.80903 & 21.00413    & 129.50018        & 302.41767        &  78.04957        & 38.54802 &
                      387.69666 &  3068.90036 &    101.85733      &  $-$27.33410      \\
             Q$_{1}$            & 163.59554 & 0.77236 & 15.17333    &  84.98641        &  41.12035        &  27.43624        & 35.30101 &
                      292.78879 &  2091.24838 &     31.32098      &  $-$46.88456      \\
             Q$_{3}$            & 319.66104 & 0.85096 & 24.42556    & 177.62272        & 319.26614        & 191.85773        & 44.92250 &
                      600.27976 &  5711.93971 &    134.13647      &      7.87076      \\
             IQR                & 156.06550 & 0.07860 &  9.25224    &  92.63631        & 278.14580        & 164.42150        &  9.62149 &
                      307.49097 &  3620.69133 &    102.81548      &     54.75532      \\
             OL                 & $-$70.50270 & 0.65446 &  1.29497 & $-$53.96806       &$-$376.09834      &$-$219.19601      & 20.86879 &
                   $-$168.44767 & $-$3339.78861 & $-$122.90224    & $-$129.01753      \\
             OU                 & 553.75928 & 0.96886 & 38.30391 & 316.57718           & 736.48484        & 438.48998        & 59.35473 &
                     1061.51622 & 11142.97670 &  288.35969        &   90.00373        \\
          \hline
        \end{tabular}
        \label{bary}
      \end{table*}
%
%

  \section{Commensurability maps}
     In this section we construct commensurability maps by generating pairs of virtual ETNOs from the data in Tables \ref{helio} and 
     \ref{bary} via Monte Carlo techniques (Metropolis \& Ulam 1949; Press et al. 2007) and evaluating their commensurability parameter or 
     ratio of orbital periods, $(a_{j}/a_{i})^{3/2}$, enforcing that $a_{j} > a_{i}$ by swapping data if necessary. The value of the 
     semimajor axis of a virtual ETNO is computed using the expression $a_{\rm v}=\langle{a}\rangle+\sigma_{a}\,r_{\rm i}$, where $a_{\rm 
     v}$ is the semimajor axis of the virtual ETNO, $\langle{a}\rangle$ is the mean value of the semimajor axis from the available orbit 
     (Tables \ref{helio} and \ref{bary}), $\sigma_{a}$ is the standard deviation of $a$ (Table \ref{helio}), and $r_{\rm i}$ is a (pseudo) 
     random number with normal distribution. In our calculations, the Box-Muller method (Box \& Muller 1958; Press et al. 2007) was used to 
     generate random numbers from the standard normal distribution with mean 0 and standard deviation 1. When computers are used to produce 
     a uniform random variable (to seed the Box-Muller method) it will inevitably have some inaccuracies because there is a lower bound on 
     how close numbers can be to 0. For a 64 bits computer the smallest non-zero number is $2^{-64}$ which means that the Box-Muller method 
     will not produce random variables more than 9.42 standard deviations from the mean; however, unphysical values have not been used in 
     the calculations. The number of pairs of virtual ETNOs tested per Monte Carlo experiment is $n=10^6$.

     \subsection{Uniform and biased expectations}
        A random (uniform) distribution of semimajor axes should exhibit a degree of commensurability compatible with zero. We have 
        generated ten sets of 16 values of the semimajor axis in the range that appears in Table \ref{helio} or \ref{bary} and assigned the
        uncertainties in Table \ref{helio} to them. We have obtained the frequency distribution of commensurability parameters for each set
        and computed the average frequency distribution. The results appear in purple in Figs \ref{commhelio} and \ref{commbary}. However,
        the ETNOs have been discovered from the ground and this fact induces a bias in the observed distribution of semimajor axes even if
        the ETNOs are not subjected to any external perturbation and just follow heliocentric, or barycentric, orbits. Following the 
        analysis in de la Fuente Marcos \& de la Fuente Marcos (2014) and, in order to reproduce this bias, we generate ten sets of 16 values 
        of the semimajor axis according to the distribution in the top panel of fig. 3 in de la Fuente Marcos \& de la Fuente Marcos (2014);
        i.e. objects with perigee $<90$~au and declination at perigee in the range ($-24\degr, 24\degr$), see fig. 2 in de la Fuente Marcos 
        \& de la Fuente Marcos (2014). The average frequency distribution of this biased sample appears in green in Figs \ref{commhelio} and 
        \ref{commbary}. The error bars have been computed assuming Poisson statistics, $\sigma=\sqrt{n}$, and using the approximation given 
        by Gehrels (1986) when $n<21$: $\sigma \sim 1 + \sqrt{0.75 + n}$, where $n$ is the number of pairs. In both cases the results are 
        compatible with a null degree of commensurability. Any statistically significant deviation from this null result should be 
        interpreted as resulting from a present-day gravitational perturbation located within the region travelled by the ETNOs.  

     \subsection{Heliocentric orbits}
\begin{figure}
        \centering
         \includegraphics[width=\linewidth]{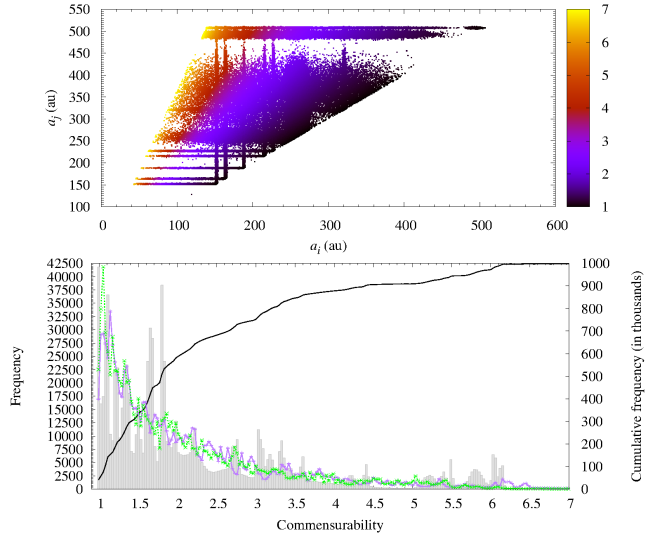}
         \caption{Commensurability map (top panel) for the values of the heliocentric semimajor axes and their uncertainties in Table 
                  \ref{helio}. Frequency distribution (bottom panel) in $(a_{j}/a_{i})^{3/2}$ from the commensurability map. The results 
                  from a uniform spread in semimajor axis are plotted in purple and those from a biased uniform sample in green. The number 
                  of bins in the frequency distribution plot is 2 $n^{1/3}$, where $n$ is the number of pairs of virtual ETNOs tested, 
                  $n=10^6$. The black curve shows the cumulative distribution. 
                 }
         \label{commhelio}
      \end{figure}

        Using the data in Table \ref{helio}, we created $10^6$ pairs of virtual ETNOs following the procedure outlined above and computed 
        the commensurability parameter. The commensurability map in terms of colour is shown in Fig. \ref{commhelio} (the value of the 
        commensurability parameter or ratio of orbital periods is plotted in colour, top panel), where each point represents a pair of 
        virtual ETNOs; the associated frequency distribution is plotted in Fig. \ref{commhelio} (bottom panel). There are obvious and 
        statistically significant deviations from the uniform or biased expectations explored above. The most conspicuous commensurabilities 
        in Fig. \ref{commhelio} (bottom panel) are 1.0, 1.11, 1.65 and 1.8. 
%
%
%
%

     \subsection{Barycentric orbits}
\begin{figure}
        \centering
         \includegraphics[width=\linewidth]{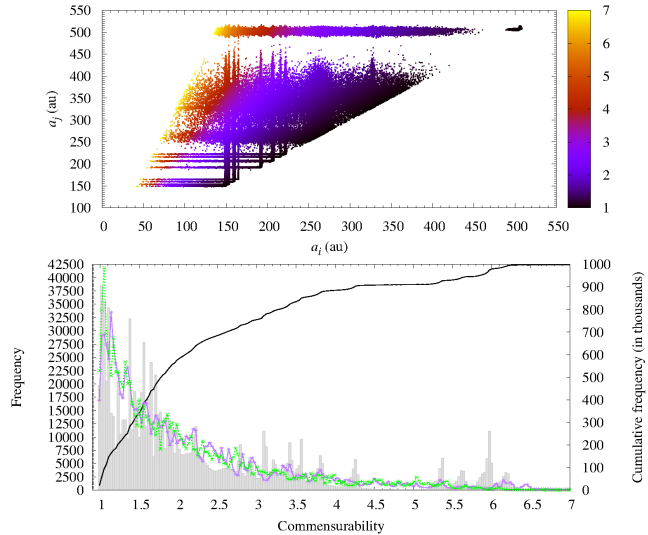}
         \caption{Same as Fig. \ref{commhelio} but for the barycentric values in Table \ref{bary}. 
                 }
         \label{commbary}
      \end{figure}

        Using the data in Table \ref{bary}, we have obtained Fig. \ref{commbary}. As for the heliocentric orbits, there are significant 
        deviations from the unperturbed scenario. Considering the values of the barycentric semimajor axes, the most obvious 
        commensurabilities in Fig. \ref{commbary} are 1.02, 1.11, 1.38, 1.65 and 1.71. Three of the statistically significant 
        commensurabilities coincide with those found in the analysis of heliocentric orbits. In addition, several of these values are also 
        present in the sequences associated with the main and trans-Neptunian belts. This is unlikely to be the result of chance alone. 
%
%
%
%

  \section{Discussion}
\begin{figure}
        \centering
         \includegraphics[width=\linewidth]{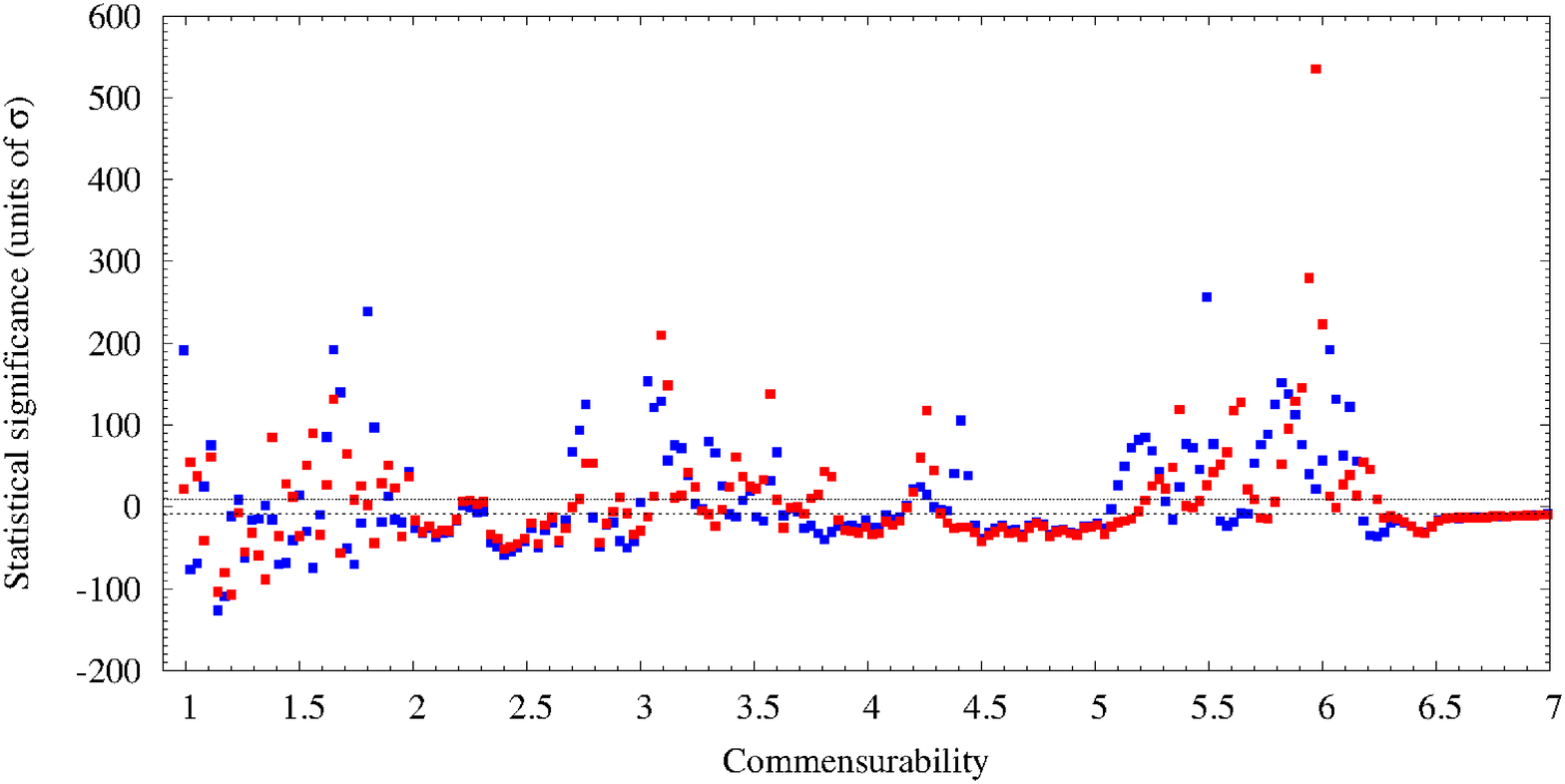}
         \caption{Statistical significance in terms of $\sigma$ of the deviations of the frequency distribution in the heliocentric (blue) 
                  and barycentric (red) cases (data from Figs \ref{commhelio} and \ref{commbary}) with respect to the biased expectations. 
                  The $\pm9\sigma$ limits are also indicated.
                 }
         \label{sigma}
      \end{figure}

     The analysis presented in the previous section shows that the frequency distribution of the ratio of orbital periods of known ETNOs is 
     statistically incompatible with that of an unperturbed asteroid population following heliocentric/barycentric orbits. This result is
     unlikely to be a statistical artefact as the known ETNOs have been discovered by several independent surveys with, presumably, 
     uncorrelated biases. If the frequency distribution of the ratio of periods is incompatible with an unperturbed scenario, how 
     significant are the favoured commensurabilities? In order to answer this rather critical question, we have computed the difference
     between the frequency obtained from the observational values and that from the biased expectation (green curve in Figs \ref{commhelio}
     and \ref{commbary}) and divided by the value of the standard deviation in the biased case calculated as described above. Our results 
     are plotted in Fig. \ref{sigma}. The number of statistically significant commensurabilities is larger than the few indicated above and
     found by simple visual inspection. The outcome is not dependent on the bin size; we have experimented with $n=10^4$ and $n=10^5$ and 
     the results are fully consistent. The statistical significance analysis in Fig. \ref{sigma} shows that our results are robust. 
     Probabilistically speaking, there are simply too many ETNOs in commensurabilities compatible with the action of massive perturbers.
     Regarding the extremely significant value of the ratio of periods $\sim6$ in Fig. \ref{sigma}, we would like to downplay this issue as
     two objects have values of the semimajor axis of $\sim500$~au and several others have semimajor axes close to 150~au. These two values
     are the abrupt cutoffs in the distribution of semimajor axes.

%
%
%
%
     One of the most surprising properties of the sample studied is the unusually large fraction of objects with ratio of periods $\sim1$.
     This issue was already noticed in de la Fuente Marcos \& de la Fuente Marcos (2014). In particular, the orbits of 2004~VN$_{112}$ and
     2013~RF$_{98}$ are alike, and several objects have values of the barycentric semimajor axis within narrow ranges of each other. This
     could mean that some of these objects move co-orbital, i.e in a 1:1 mean motion resonance, with an unseen planet (or, more likely, 
     several of them). Co-orbital motion is possible at high eccentricity (Namouni 1999; Namouni, Christou \& Murray 1999; Namouni \& Murray 
     2000). However, a more likely scenario is in the presence of higher order resonances like the 7:3 and 5:2 in the main asteroid belt 
     that give a ratio of periods of 1.0699 or the 4:7 and 3:5 in the trans-Neptunian belt that give a ratio of 1.0501. Interior and 
     exterior resonances are possible if there are multiple trans-Plutonian planets. Regarding the impact of this resonant scenario on the 
     Planet Nine hypothesis, it is clearly compatible with it.

     If the mechanisms responsible for inducing dynamical structure in the main and trans-Neptunian belts are also at work in the region
     occupied by the ETNOs then we should expect that for two given ETNOs in near commensurability:
     \begin{equation}
        \left(\frac{a_{\rm p}}{a_{j}}\right)^{3/2} \ \left(\frac{a_{i}}{a_{\rm p}}\right)^{3/2} 
             = \left(\frac{m}{n}\right) \ \left(\frac{k}{l}\right) \,, \label{key}
     \end{equation}
     where $a_{\rm p}$ is the semimajor axis of the orbit of the perturber and $k, l, m, n$ are all small integers. This expression can be
     applied to two of the most clear clusterings in semimajor axis in Table \ref{bary}. The average value of the barycentric semimajor 
     axis of (90377) Sedna and 2007 TG$_{422}$ is 504~au. On the other hand, the equivalent mean value for 2004 VN$_{112}$, 2010 GB$_{174}$
     and 2013~RF$_{98}$ is 332~au. These five objects are part of the set of six singled out by Batygin \& Brown (2016). The associated 
     period ratio for these two sets of ETNOs is 1.87. In the main asteroid belt, this ratio is obtained for objects trapped in the 5:3
     mean motion resonance with Jupiter and those in the 3:1, that is one of the main resonances in the outer belt (Holman \& Murray 1996).
     Making a dynamical analogy between the two situations and decomposing Eqn. \ref{key} in two we have: $(a_{\rm p}/504)^{3/2}=5/3$ and 
     $(332/a_{\rm p})^{3/2}=1/3$. The average of the two values of $a_{\rm p}$ is $\sim$700~au which is the favoured value for the 
     semimajor axis of Planet Nine in Batygin \& Brown (2016). The 1.8 commensurability has a statistical significance of 239$\sigma$ for 
     heliocentric orbits, the 1.89 commensurability has 51$\sigma$ for barycentric orbits (see Fig. \ref{sigma}). This is unlikely to be 
     mere coincidence.

     Another example of the potential implications of our findings arises when we focus on 2003 HB$_{57}$, 2015 SO$_{20}$, 2005 RH$_{52}$, 
     (445473) 2010 VZ$_{98}$ and 2013~GP$_{136}$, the first two could be in a 3:2 resonance with a hypothetical planet at $a=213$~au, with 
     the other three in a 5:3 resonance with the same planet. In this framework, the pair 2003 HB$_{57}$ and 2015 SO$_{20}$ would be in a 
     10:9 accidental resonance with the other three ETNOs. A ratio of periods $\sim1.1$ is present in both the main asteroid belt and the 
     trans-Neptunian belt. The 1.1 commensurability has a statistical significance of 76$\sigma$ for heliocentric orbits and 61$\sigma$ for 
     barycentric orbits (see Fig. \ref{sigma}). On the other hand, the 1.65 commensurability is present for both heliocentric (192$\sigma$)
     and barycentric (131$\sigma$) orbits; a similar analysis focusing on 2003 HB$_{57}$, 2013~GP$_{136}$, (82158) 2001 FP$_{185}$ and 2002 
     GB$_{32}$ is compatible with a hypothetical planet at $a=329$~au considering resonances (3/1)(5/9)=5/3$\sim$1.66. With the currently 
     available data, degenerate solutions are possible, but they still hint at a multi-planet scenario. Trans-Plutonian planets may have 
     been scattered out of the region of the giant planets early in the history of the Solar system (see e.g. Bromley \& Kenyon 2014, 2016), 
     but planets similar to Uranus or Neptune may also form at 125--750 au from the Sun (Kenyon \& Bromley 2015, 2016). 

  \section{Conclusions}
     In this Letter, we have explored the existence of commensurabilities between the known ETNOs. This analysis has been inspired by the 
     hypothesis and discussion presented in Malhotra et al. (2016). Given the fact that these objects are not massive enough to induce 
     orbit-orbit coupling by themselves, the existence of statistically significant commensurabilities could only signal the presence of 
     massive unseen perturbers. Summarizing, our conclusions are as following.
     \begin{itemize}
        \item The clustering in orbital parameter space observed for the heliocentric orbits of the known ETNOs is also present when 
              considering their barycentric orbits. In particular, $e$ clumps about 0.81$\pm$0.06, $i$ about 20\degr$\pm$8\degr, $\Omega$ 
              about 134\degr$\pm$72\degr, and $\omega$ about $-$26\degr$\pm$49\degr. Statistical outliers do exist.
        \item The frequency distribution of the ratio of orbital periods of known ETNOs is statistically incompatible with that of an
              unperturbed asteroid population following heliocentric/barycentric orbits. In contrast, it resembles the ones present in the 
              main and trans-Neptunian belts. The existence of statistically significant accidental commensurabilities between ETNOs 
              strongly suggests that external perturbers induce the observed dynamical structure.
        \item The fraction of known ETNOs in 1:1 commensurabilities is dozens of times the expected value for a random population.
        \item A number of known ETNOs may be trapped in the 5:3 and 3:1 mean motion resonances with a putative Planet Nine with semimajor 
              axis $\sim$700~au. 
     \end{itemize}
     We must stress that our results are based on small number statistics, but it is also true that the deviations from what is expected
     for an unperturbed asteroid population following heliocentric/barycentric orbits are so strong that what is observed is unlikely to
     be the result of chance alone. Gravitational perturbations with sources in the region travelled by the ETNOs are the most probable
     explanation for the observed patterns.  

  \section*{Acknowledgements}
     We thank the anonymous referee for her/his constructive and helpful report, and S. J. Aarseth, D. P. Whitmire, G. Carraro, D. Fabrycky, 
     A. V. Tutukov, S. Mashchenko, S. Deen and J. Higley for comments on ETNOs and trans-Plutonian planets. This work was partially 
     supported by the Spanish `Comunidad de Madrid' under grant CAM S2009/ESP-1496. In preparation of this Letter, we made use of the NASA 
     Astrophysics Data System, the ASTRO-PH e-print server, and the MPC data server.

  \bsp
  \label{lastpage}

\begin{thebibliography}{99}
     \bibitem[\protect\citeauthoryear{Batygin \& Brown}{2016}]{BB16} Batygin K., Brown M. E., 2016,
             AJ, 151, 22
     \bibitem[\protect\citeauthoryear{Box \& Muller}{1958}]{BM58} Box G. E. P., Muller M. E., 1958,
             Annals Math. Stat., 29, 610
     \bibitem[\protect\citeauthoryear{Bromley \& Kenyon}{2014}]{BK14} Bromley B. C., Kenyon S. J., 2014,
             ApJ, 796, 141
     \bibitem[\protect\citeauthoryear{Bromley \& Kenyon}{2016}]{BK16} Bromley B. C., Kenyon S. J., 2016,
             ApJ, submitted (arXiv:1603.08010)
     \bibitem[\protect\citeauthoryear{de la Fuente Marcos}{2014}]{FM14} de la Fuente Marcos C., de la Fuente Marcos R., 2014,
             MNRAS, 443, L59
     \bibitem[\protect\citeauthoryear{de la Fuente Marcos}{2016}]{FM16} de la Fuente Marcos C., de la Fuente Marcos R., 2016,
             MNRAS, 459, L66
     \bibitem[\protect\citeauthoryear{de la Fuente Marcos et al.}{2015}]{FM15} de la Fuente Marcos C., de la Fuente Marcos R., Aarseth S. J., 2015,
             MNRAS, 446, 1867
     \bibitem[\protect\citeauthoryear{Gehrels}{1986}]{GE86} Gehrels N., 1986,
             ApJ, 303, 336
     \bibitem[\protect\citeauthoryear{Giorgini}{1996}]{GI96} Giorgini J. D. et al., 1996,
             BAAS, 28, 1158
     \bibitem[\protect\citeauthoryear{Gladman et al.}{2012}]{GL12} Gladman B. et al., 2012,
             AJ, 144, 23
     \bibitem[\protect\citeauthoryear{Gomes et al.}{2005}]{GO05} Gomes R. S., Gallardo T., Fern\'andez J. A., Brunini A., 2005,
             Celest. Mech. Dyn. Astron., 91, 109
     \bibitem[\protect\citeauthoryear{Gomes et al.}{2015}]{GO15} Gomes R. S., Soares J. S., Brasser R., 2015,
             Icarus, 258, 37
     \bibitem[\protect\citeauthoryear{Holman \& Murray}{1996}]{HM96} Holman M. J., Murray N. W., 1996,
             AJ, 112, 1278
     \bibitem[\protect\citeauthoryear{Kenyon \& Bromley}{2015}]{BK15} Kenyon S. J., Bromley B. C., 2015,
             ApJ, 806, 42
     \bibitem[\protect\citeauthoryear{Kenyon \& Bromley}{2016}]{KB16} Kenyon S. J., Bromley B. C., 2016,
             ApJ, submitted (arXiv:1603.08008)
     \bibitem[\protect\citeauthoryear{Malhotra et al.}{2016}]{MA16} Malhotra R., Volk K., Wang X., 2016,
             ApJ, submitted  (arXiv:1603.02196)
     \bibitem[\protect\citeauthoryear{Metropolis}{1949}]{MU49} Metropolis N., Ulam S., 1949,
             J. Am. Stat. Assoc., 44, 335
     \bibitem[\protect\citeauthoryear{Murray \& Dermott}{1999}]{MU99} Murray C. D., Dermott S. F., 1999,
             Solar system Dynamics.
             Cambridge Univ. Press, Cambridge, p.\ 97
     \bibitem[\protect\citeauthoryear{Namouni}{1999}]{NA99} Namouni F., 1999,
             Icarus, 137, 293
     \bibitem[\protect\citeauthoryear{Namouni \& Murray}{2000}]{NA00} Namouni F., Murray C. D., 2000,
             Celest. Mech. Dyn. Astron., 76, 131
     \bibitem[\protect\citeauthoryear{Namouni et al.}{1999}]{NC99} Namouni F., Christou A. A., Murray C. D., 1999,
             Phys. Rev. Lett., 83, 2506
     \bibitem[\protect\citeauthoryear{Nesvorn\'y \& Morbidelli}{1999}]{NM99} Nesvorn\'y D., Morbidelli A., 1999,
             in Henrard J., Ferraz-Mello S., eds,
             Proc. IAU Colloq. 172, Impact of Modern Dynamics in Astronomy. 
             Kluwer Academic Publishers, Dordrecht, p.\ 381
     \bibitem[\protect\citeauthoryear{Press}{2007}]{PR07} Press W. H., Teukolsky S. A., Vetterling W. T., Flannery B. P., 2007,
             Numerical Recipes: The Art of Scientific Computing, 3rd edn.
             Cambridge Univ. Press, Cambridge
     \bibitem[\protect\citeauthoryear{Todorovic-Juchnicwicz}{1981}]{TO81} Todorovic-Juchnicwicz B., 1981,
             Acta Astron., 31, 191
     \bibitem[\protect\citeauthoryear{Trujillo \& Sheppard}{2014}]{TS14} Trujillo C. A., Sheppard S. S., 2014,
             Nature, 507, 471
  \end{thebibliography}
\end{document}